\documentclass[preprint, 10pt, 5p]{elsarticle}
\usepackage[utf8]{inputenc}
\usepackage{amsmath}
\usepackage{amsfonts}
\usepackage{amssymb}
\usepackage{mathrsfs}
\usepackage{subfigure}
\usepackage[hidelinks]{hyperref}
\usepackage{outlines}
\usepackage{tikz}
\usepackage{pgfplots}

\biboptions{numbers, sort&compress}

\newcommand{\D}{\mathrm{d}}

\usepackage{tensor}

\newcommand{\beq}{\begin{equation}}
\newcommand{\eeq}{\end{equation}}

\newcommand{\bvec}{\begin{pmatrix}}
\newcommand{\evec}{\end{pmatrix}}

\begin{document}

\title{Fluid Model for the Piezothermal Effect}
\date{\today}
\author{E. J. Kolmes}
\ead{ekolmes@princeton.edu}
%\author{without email, }
\address{Department of Astrophysical Sciences, Princeton University, Princeton, NJ 08544, USA}
\author{V. I. Geyko}
\ead{geyko1@llnl.gov}
\address{Lawrence Livermore National Laboratory, Livermore, California, 94550, USA}
\author{N. J. Fisch}
\ead{fisch@princeton.edu}
\address{Department of Astrophysical Sciences, Princeton University, Princeton, NJ 08544, USA}

\begin{abstract}
When a gas in an externally imposed potential field is compressed, temperature gradients appear. This has been called the piezothermal effect. It is possible to analytically calculate the time-dependent behavior of the piezothermal effect using a linearized fluid model. Quantitative differences between the fluid-model results and previous numerical calculations can be explained by the effects of viscosity and heat conductivity. The fluid model casts the piezothermal effect as a spectrum of buoyancy oscillations, which yields new physical insights into the effect.  
\end{abstract}

\begin{keyword}
Rotating fluid, piezothermal effect, compression, Brunt-V\"ais\"al\"a oscillations
\end{keyword}

\maketitle

\section{Introduction}

Consider a gas at rest in a potential field. If the gas is compressed, it will be heated. Moreover -- contrary to the usual intuition about compressional heating -- the resulting temperature will be spatially nonuniform, such that regions that are higher in the potential well are hotter. This effect was described by Geyko and Fisch \cite{Geyko2016} and called the \textit{piezothermal} effect. Intuitively, it results from the fact that particles starting in equilibrium move toward (and further compress) regions of higher potential as they are heated. 

In the original paper on the piezothermal effect, Geyko and Fisch observed the phenomenon in particle simulations. Analytically, they used a toy model to explain the scalings and some of the quantitative behavior of the simulations. Their model described the gas as two homogeneous regions separated by a massive movable membrane, so that the two sides of the system could have different temperatures and densities and could exert pressure on one another. For the simulation tools, they used a one-dimensional Monte Carlo code with exact energy and momentum conservation properties and a hard-sphere binary-collision operator. While their models correctly described the essential characteristics of the effect, they left room for discussion and future improvement in a number of respects. 
%While in general they obtained quite adequate results, there were still a room for discussion and future improvement, which is done in the present work.

This paper analyzes the piezothermal effect by instead using a fluid model. 
The fluid approach to the piezothermal effect makes it possible to analytically calculate the behavior of the piezothermal effect in a wider range of scenarios, in greater detail, and using fewer simplifying assumptions than was done previously. Numerical fluid simulations confirm the validity of the analytic model and -- when compared in detail to the results of the Monte Carlo code used in the original paper -- help to explain quantitative discrepancies between the fluid-model results and the previous numerical results. 

The piezothermal effect is closely related to the physics to the rotation-dependent heat capacity effect also studied by Geyko and Fisch, in which the energy required to compress a rotating cylinder changes when the gas is spinning \cite{Geyko2013, Geyko2017}. That effect has applications in engine design, where it could be used to improve the efficiency of Otto and Diesel cycles \cite{Geyko2014}. 
In addition, the piezothermal effect is phenomenologically similar to the behavior observed in Ranque-Hilsch vortex tubes, which also produce radial temperature gradients in a rotating gas \cite{Ranque1933, Hilsch1947, Kassner1948, Ahlborn1997, Ahlborn1998, Ahlborn2000, Liew2012, Kolmes2017}. Vortex tubes are used for spot cooling in a variety of industrial applications. 
In general, the ability to move energy in rotating and compressing systems -- either spatially or between degrees of freedom -- can be of great practical utility \cite{Geyko2014, Davidovits2016}. 
These effects can also be useful for understanding the natural world. In particular, the fluid treatment of the piezothermal effect makes it clear that there is a strong connection between the piezothermal effect and Brunt-V\"ais\"al\"a oscillations, which are observed in a variety of naturally stratified media \cite{Brunt1927, Durran1982, Emery1984, Brassard1991}.

\section{Linearized Fluid Model for Fast Compression} \label{sec:fastCompression}

\begin{figure}
	\centering
		\begin{tikzpicture} [
		declare function={a(\x)=\x;},
		declare function={b(\x)=0.5*\x-1;}
		]
		
		\draw[very thick, black] (0,0) rectangle (4,7);
		
		\draw[->, very thick, blue] (4.5,6.0) -- (4.5,5.0);
		\node at (4.8, 5.6) {\Large \color{blue} $g$}; 
		
		\draw[->, very thick, black] (-2.25,3.5) -- (-.2,3.5); 
		\draw[->, very thick, black] (6.25,3.5) -- (4.2,3.5);
		\node at (-1.3, 4.0) {compression};
		
		\draw[->, very thick, red] (-0.5, 5.0) -- (-0.5, 6.0); 
		\node at (-1.1, 5.6) {\Large \color{red} $\nabla T$}; 
		
		\fill[olive] (1.0,1.1) circle (.1); 
		\fill[olive] (2.1,0.8) circle (.1); 
		\fill[olive] (3.4,1.3) circle (.1);
		\fill[olive] (0.4,0.5) circle (.1); 
		\fill[olive] (3.2,3.3) circle (.1); 
		\fill[olive] (1.9,2.8) circle (.1); 
		\fill[olive] (0.9,3.1) circle (.1); 
		\fill[olive] (1.4,5.0) circle (.1); 
		\fill[olive] (0.3,1.7) circle (.1); 
		\fill[olive] (3.7,1.8) circle (.1); 
		\fill[olive] (3.2,4.4) circle (.1);
		\fill[olive] (2.8,2.3) circle (.1);
		\fill[olive] (3.5,0.4) circle (.1);
		\fill[olive] (1.6,1.6) circle (.1);
		\fill[olive] (0.6,5.9) circle (.1);
		\fill[olive] (2.9,6.2) circle (.1);
		\fill[olive] (3.0,0.8) circle (.1);
		\fill[olive] (1.3,0.6) circle (.1);
		\fill[olive] (2.5,0.4) circle (.1);
		\fill[olive] (2.2,3.8) circle (.1);
		\fill[olive] (0.7,4.2) circle (.1);
		\fill[olive] (1.1,2.1) circle (.1);
		\fill[olive] (2.4,1.9) circle (.1);
		
%		\begin{axis}[
%				xmin = 0, xmax = 1, 
%				ymin = 0, ymax = 1,
%		domain=0:4,
%		xmin = 0,
%		axis lines=none,
%		axis equal image,
%		xtick=\empty, ytick=\empty,
%		enlargelimits=true,
%		clip mode=individual, clip=false
%		]
%		\addplot [olive, only marks, mark=*, samples=100, mark size=2]
%				{0.5*(a(x)+b(x)) + 0.5*rand*(a(x)-b(x))};
%		{rand}; 
%		\end{axis}
		
		\end{tikzpicture}
	\caption{This schematic shows a simple setup that demonstrates the piezothermal effect. Compression transverse to the direction of gravity produces temperature gradients parallel with gravity and in the opposite direction. } \label{fig:compressionCartoon}
\end{figure}
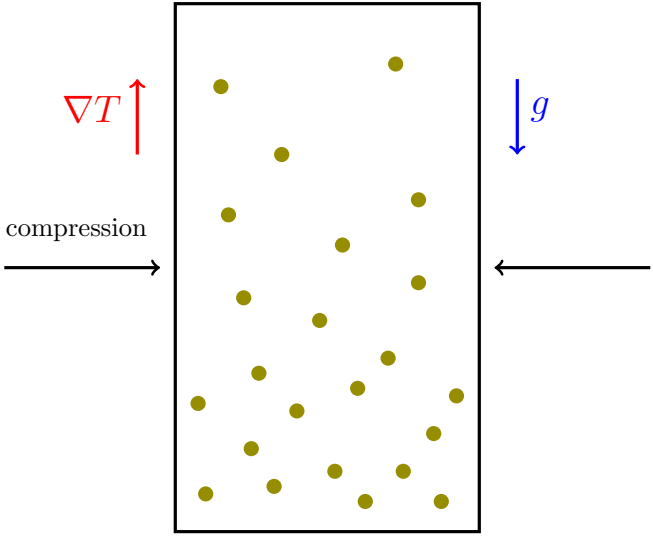

For simplicity, we consider the potential field to be gravitational, although practical applications are more likely in spinning systems, where centrifugal forces take the role of gravitational forces. Thus, to describe the key effects most simply, consider a gas in a gravitational field, such that all quantities vary only in the direction of the field. Suppose the fluid is compressed in a direction perpendicular to the gravitational field. The behavior of the system depends on four timescales: the collisional timescale $\tau_c$, the compression timescale $\tau_E$, the sound timescale $\tau_s$, and the timescale $\tau_H$ associated with spatial heat conduction. Geyko and Fisch studied the piezothermal effect in a fast-compression scenario and in a slow-compression scenario. In the fast-compression scenario, $\tau_c \ll \tau_E \ll \tau_s \ll \tau_H$. The first part of this inequality implies that the gas is always in local equilibrium. The second inequality means that the input of energy due to compression happens much more quickly than the system can react spatially. The last part of the inequality states that spatial heat conductivity can be neglected. 

Because of the very fast collisional timescale, it is appropriate to describe the system with a fluid model (a system with less frequent collisions could behave very differently \cite{Kolmes2016}). Using an adiabatic equation of state, the fluid density, velocity, and temperature can be modeled by 
\begin{gather}
\frac{\partial n}{\partial t} + \frac{\partial}{\partial x} \big(n v \big) = 0 \label{eqn:nonlinearN} \\
m n \bigg( \frac{\partial v}{\partial t} + v \frac{\partial v}{\partial x} \bigg) = - \frac{\partial (nT)}{\partial x} - m n g \label{eqn:nonlinearV} \\
\bigg( \frac{\partial}{\partial t} + v \frac{\partial}{\partial x} \bigg)\bigg( \frac{T}{n^{\gamma-1}} \bigg) = 0. \label{eqn:nonlinearT}
\end{gather}
Suppose the system is bounded between $x = 0$ and $x = L$. Define equilibrium profiles 
\begin{gather}
n_0(x) \doteq \bigg(\frac{mg/T_0}{1-e^{-mgL/T_0}}\bigg) \, e^{-mgx/T_0} \\
T_0(x) \doteq T_0 = \text{const} \\
v_0(x) \doteq 0. 
\end{gather}
Now suppose the system is perturbed so that at $t = 0$, the temperature is (uniformly) changed from $T_0$ to $T_i$. This can occur, for example, by lateral compression as shown in Figure~\ref{fig:compressionCartoon}. Define 
\begin{gather}
\delta \doteq \frac{T_i - T_0}{T_0} 
\end{gather}
and suppose $\delta \ll 1$. $n$, $T$, and $v$ can be expanded about equilibrium so that 
\begin{gather}
n = n_0 + n_1 + \mathcal{O}(\delta^2) \\
T = T_0 + T_1 + \mathcal{O}(\delta^2) \\
v = v_1 + \mathcal{O}(\delta^2) . 
\end{gather}
The initial conditions for $n_1$, $T_1$, and $v_1$ are 
\begin{gather}
n_1 \big|_{t=0} = 0 \label{eqn:initialN1} \\
T_1 \big|_{t=0} = T_0 \delta \label{eqn:initialT1} \\
v_1 \big|_{t=0} = 0. \label{eqn:initialV1}
\end{gather}
The initial conditions for their time derivatives can be derived by combining these with the equations of motion. 
Define the equilibrium scale height $z_0$ by 
\begin{gather}
z_0 \doteq \frac{T_0}{mg} \, .
\end{gather}To first order in $\delta$, the equations of motion can be written as 
\begin{gather}
\frac{\partial n_1}{\partial t} = \frac{1}{z_0} \, n_0 v_1 - n_0 \frac{\partial v_1}{\partial x} \label{eqn:nLinear} \\
\frac{\partial v_1}{\partial t} = \frac{1}{m z_0} \, T_1 - \frac{1}{m} \frac{\partial T_1}{\partial x} - \frac{T_0}{m n_0} \frac{\partial n_1}{\partial x} - \frac{T_0}{m z_0 n_0} \, n_1 \label{eqn:vLinear} \\
\frac{\partial T_1}{\partial t} = (\gamma-1) \frac{T_0}{n_0} \bigg( \frac{\partial n_1}{\partial t} - \frac{1}{z_0} \, v_1 n_0 \bigg) \label{eqn:TLinear} .
\end{gather}
Taking an additional time derivative of Eq.~(\ref{eqn:vLinear}) and plugging in Eqs.~(\ref{eqn:nLinear}) and (\ref{eqn:TLinear}),
\begin{gather}
\frac{\partial^2 v_1}{\partial t^2} = \frac{\gamma T_0}{m} \bigg( \frac{\partial^2 v_1}{\partial x^2} -  \frac{1}{z_0} \frac{\partial v_1}{\partial x} \bigg) .
\end{gather}
Define $c_s^2 \doteq \gamma T_0 / m$ and $f \doteq v_1 e^{-x/2z_0}$. Then 
\begin{gather}
\frac{\partial^2 f}{\partial t^2} = c_s^2 \bigg( \frac{\partial^2 f}{\partial x^2} - \frac{1}{4 z_0^2} \, f \bigg) . \label{eqn:fWave}
\end{gather}
Applying the boundary conditions at $x = 0$ and $x = L$, $f$ can be written as 
\begin{gather}
f(t,x) = \sum_{n=1}^\infty \Xi_n(t) \sin \bigg( \frac{\pi n x}{L} \bigg) \, 
\end{gather} 
for some functions $\Xi_n(t)$. Then Eq.~(\ref{eqn:fWave}) implies 
\begin{gather}
\ddot{\Xi}_n(t) = - c_s^2 \bigg( \frac{\pi^2 n^2}{L^2} + \frac{1}{4 z_0^2} \bigg) \Xi_n. 
\end{gather}
The time-dependent coefficients are linear combinations of sine and cosines in time. In order to get $v_1 = 0$ at $t=0$, only the sine terms can survive. As such, 
\begin{gather}
f(t,x) = \sum_{n=1}^\infty \alpha_n \sin (k_n x) \sin (\omega_n t) \label{eqn:fSeriesUndetermined}
\end{gather}
for some constants $\alpha_n$, with $k_n$ and $\omega_n$ defined by 
\begin{gather}
k_n \doteq \frac{\pi n}{L} \\
\omega_n \doteq c_s \sqrt{ \frac{\pi^2 n^2}{L^2} + \frac{1}{4 z_0^2} } = \omega_0 \sqrt{1 + 4 z_0^2 k_n^2} \, . \label{eqn:omegaN}
\end{gather}
Here $\omega_0 = c_s / 2 z_0$. 
In order to determine the constants $\alpha_n$, consider the initial condition on $\partial v_1 / \partial t$. Combining Eq.~(\ref{eqn:vLinear}) with Eqs.~(\ref{eqn:initialN1}), (\ref{eqn:initialT1}), and (\ref{eqn:initialV1}),
\begin{gather}
\frac{\partial v_1}{\partial t} \bigg|_{t=0} = g \delta, 
\end{gather}
so 
\begin{gather}
\frac{\partial f}{\partial t} \bigg|_{t=0} = g \delta e^{-x/2z_0} .
\end{gather}
The sine series for $e^{-x/\lambda}$ is 
\begin{gather}
e^{-x/\lambda} = \sum_{n=1}^\infty \frac{2 n \pi \lambda^2}{L^2 + n^2 \pi^2 \lambda^2} \big[ 1 + (-1)^{n+1} e^{-L/\lambda} \big] \sin (k_n x). 
\end{gather}
Using this, 
\begin{align}
&\frac{\partial f}{\partial t} \bigg|_{t=0} = \frac{g z_0 \delta}{L} \nonumber \\
&\hspace{9 pt}\times \sum_{n=1}^\infty \frac{8 k_n z_0}{1 + 4 k_n^2 z_0^2} \big[ 1 + (-1)^{n+1} e^{-L/2 z_0} \big] \sin(k_n x). 
\end{align}
Eq.~(\ref{eqn:fSeriesUndetermined}) implies that 
\begin{align}
\frac{\partial f}{\partial t} \bigg|_{t=0} &= \sum_{n=1}^\infty \omega_n \alpha_n \sin(k_n x). 
\end{align}
This determines the $\alpha_n$ parameters. 
\begin{align}
&f = \frac{2 g z_0^2 \delta}{L c_s} \sum_{n=1}^\infty \bigg[ \frac{8 k_n z_0}{(1 + 4 k_n^2 z_0^2)^{3/2}} \nonumber \\
&\hspace{28 pt}\times \big[ 1 + (-1)^{n+1} e^{-L/2 z_0} \big] \sin(k_n x) \sin(\omega_n t) \bigg].
\end{align}
The governing equation for $T_1$ can be written as 
\begin{align}
\frac{\partial T_1}{\partial t} &= - (\gamma - 1) T_0 \bigg( \frac{\partial f}{\partial x} + \frac{f}{2 z_0} \bigg) e^{x/2z_0}, 
\end{align}
which is
\begin{align}
&\frac{\partial T_1}{\partial t} = - \frac{\gamma - 1}{\gamma} \frac{2 c_s T_0 \delta}{L} e^{x/2z_0} \nonumber \\
&\hspace{10 pt}\times \sum_{n=1}^\infty \bigg[ \frac{4 k_n z_0}{(1+4 k_n^2 z_0^2)^{3/2}} \big[ 1 + (-1)^{n+1} e^{-L/2z_0} \big] \nonumber \\
&\hspace{30 pt}\times \bigg( \sin(k_n x) + 2 k_n z_0 \cos (k_n x) \bigg) \sin(\omega_n t) \bigg] .
\end{align}
Integrating and applying the initial condition on $T_1$, 
\begin{align}
&\frac{T_1}{T_0} = \delta - \frac{\gamma - 1}{\gamma} \frac{4 z_0 \delta}{L} e^{x/2z_0} \nonumber \\
&\hspace{5 pt}\times \sum_{n=1}^\infty \bigg[ \frac{4 k_n z_0}{(1+4 k_n^2 z_0^2)^2} \big[ 1 + (-1)^{n+1} e^{-L/2z_0} \big] \nonumber \\
&\hspace{9 pt} \times \bigg( \sin(k_n x) + 2 z_0 k_n \cos (k_n x) \bigg) [1 - \cos(\omega_n t)] \bigg] . \label{eqn:fastCompressionSolution}
\end{align}
Define the field-strength parameter $G$ as 
\begin{gather}
G \doteq \frac{L}{z_0} = \frac{m g L}{T_0} \, .
\end{gather}
In terms of $G$, 
\begin{align}
&\frac{T_1(t,x)}{T_0} = \delta - \frac{\gamma - 1}{\gamma} \big( 8 G \delta \big) e^{(x/L) (G/2)} \nonumber \\
&\hspace{5 pt}\times \sum_{n=1}^\infty \bigg[ \frac{4 \pi n}{(G^2+4 \pi^2 n^2)^2} \big[ 1 + (-1)^{n+1} e^{-G/2} \big] \nonumber \\
&\hspace{9 pt} \times \bigg( G \sin(k_n x) + 2 \pi n \cos (k_n x) \bigg) \sin^2 \bigg( \frac{\omega_n t}{2} \bigg) \bigg] . \label{eqn:fastCompressionSolutionDimensionlessParameters}
\end{align}
Qualitatively, it is clear from Eq.~(\ref{eqn:fastCompressionSolutionDimensionlessParameters}) that the shape of $T_1(t,x)$ will depend strongly on $G$. Modes other than $n=1$ will contribute significantly when $n \lesssim G / 2 \pi$. When the $n=1$ mode is dominant, the spatial and temporal structure are simple, with a well-defined wavelength and oscillation frequency. As G increases, the spatial structure becomes progressively more complicated. 

In the weak-field $G \ll 1$ limit, Eq.~(\ref{eqn:fastCompressionSolutionDimensionlessParameters}) becomes 
\begin{align}
&\lim_{G \rightarrow 0} \frac{T_1(t,x)}{T_0} = \delta - \frac{\gamma - 1}{\gamma} \big( 4 G \delta \big) \nonumber \\
&\hspace{39 pt}\times \sum_{n=1}^\infty \frac{1 + (-1)^{n+1}}{\pi^2 n^2} \cos (k_n x) \sin^2 \bigg( \frac{\omega_n t}{2} \bigg) .
\end{align}
When $G \ll 1$ and $t = L / c_s$, $\sin^2 (\omega_n t / 2) \rightarrow 1+\mathcal{O}(G^2)$ $\forall n \in \mathbb{Z}$. Therefore, the maximal temperature difference between $x = 0$ and $x = L$ is 
\begin{align}\label{eq:kappa_limit}
\lim_{G \rightarrow 0} \frac{T_1(L/c_s, L) - T_1(L/c_s, 0)}{T_0} = \frac{\gamma-1}{\gamma} (2 G \delta). 
\end{align}
When $\gamma = 5/3$, this is $0.8 G \delta$. This is precisely the analytic result found by Geyko and Fisch in this limit. However, it disagrees with the results of their simulations, in which $\Delta T_1 / T_0 \approx 0.64 G \delta$. 

\begin{figure*}
	\centering
	\includegraphics[width=\linewidth]{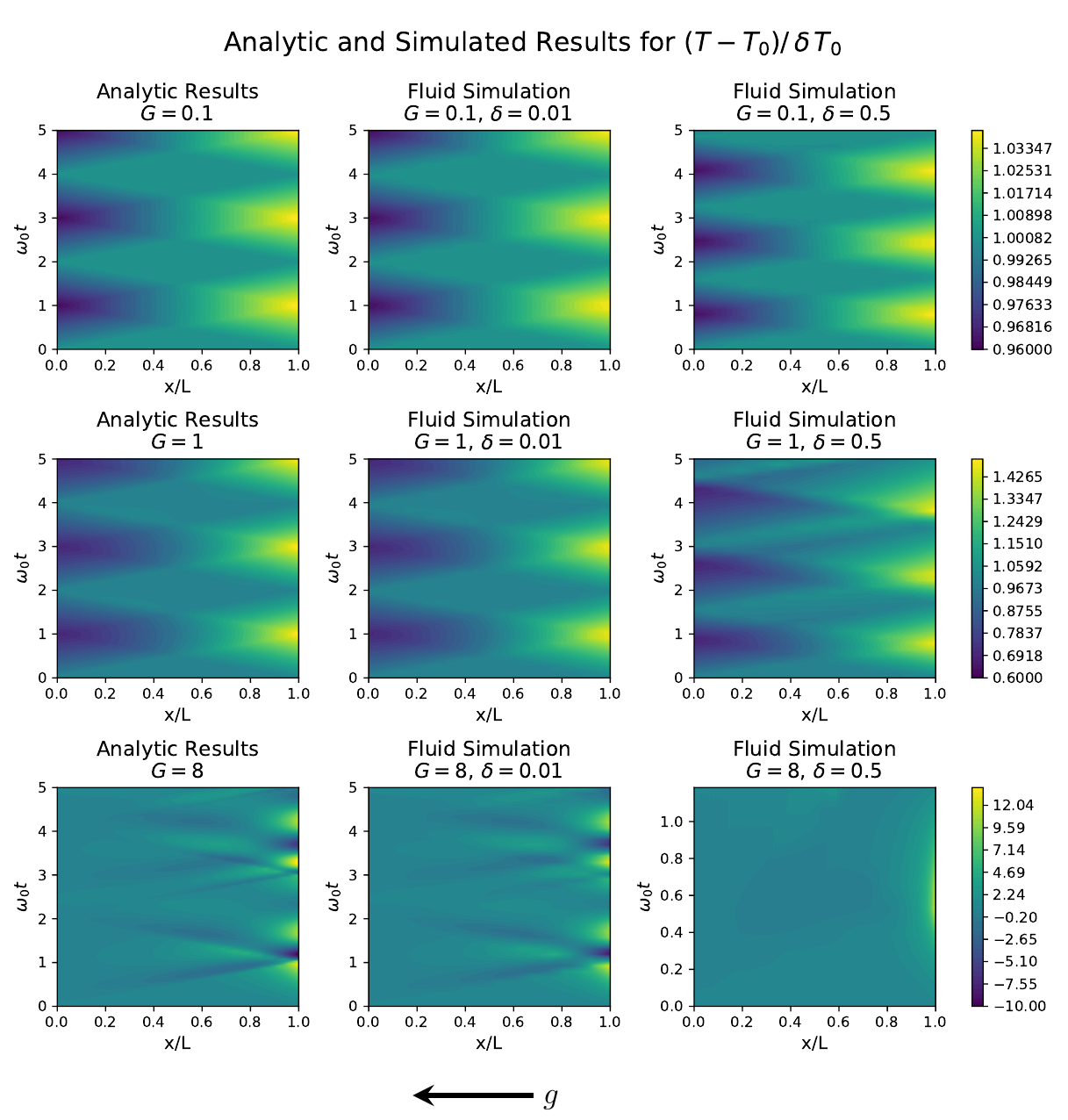}
	
	\caption{This figure shows analytic and numerical results for the temperature oscillations associated with the piezothermal effect. Each row corresponds to a different choice of $G$. The left column is the analytic result from Eq.~(\ref{eqn:fastCompressionSolutionDimensionlessParameters}). The plots in the center and on the right are numerical solutions to the full nonlinear fluid equations described by Eqs.~(\ref{eqn:nonlinearN}), (\ref{eqn:nonlinearV}), and (\ref{eqn:nonlinearT}) with $\delta = 10^{-2}$ and $\delta = 0.5$, respectively. Times are normalized to $\omega_0^{-1}$, which depends on $G$ and $T_0$. } \label{fig:simulations}
\end{figure*}

Simulations of the full nonlinear fluid equations given by Eqs.~(\ref{eqn:nonlinearN}), (\ref{eqn:nonlinearV}), and (\ref{eqn:nonlinearT}) were performed using the 1D fluid code SNeuT, which uses components of the SUNDIALS suite \cite{Hindmarsh2005, Cohen1996}. Figure~\ref{fig:simulations} shows these simulations alongside the analytically predicted results from the fluid model; when $\delta$ is small, they are in close agreement, including the coefficient of 0.8. The origin of the discrepancy between these and the original paper's results is discussed in Section \ref{sec:two_codes}. 
%However, one possibility comes from the fact that the simulations in the original paper used particle simulations with a hard-sphere collision operator and would not have been able to explicitly remove the effects of heat conductivity. It is numerically difficult to make the collisional mean free path vanish using such a scheme. In contrast, in the fluid simulations used here, the conductivity is a free parameter which can be controlled (in this context, set to zero) directly. 

Now consider the opposite limit, where $G \gg 1$: 
\begin{align}
&\lim_{G \rightarrow \infty}\frac{T_1(t,x)}{T_0} = \delta - \frac{\gamma - 1}{\gamma} \big( 8 G \delta \big) e^{(x/L) (G/2)} \nonumber \\
&\hspace{5 pt}\times \sum_{n=1}^\infty \bigg[ \frac{4 \pi n}{(G^2+4 \pi^2 n^2)^2} \bigg( G \sin(k_n x) + 2 \pi n \cos (k_n x) \bigg) \nonumber \\
&\hspace{140 pt}\times \sin^2 \bigg( \frac{\omega_n t}{2} \bigg) \bigg] .
\end{align}
This can be converted to an integral: 
\begin{align}
&\lim_{G \rightarrow \infty}\frac{T_1(t,x)}{T_0} = \delta - \frac{\gamma - 1}{\gamma} \frac{8 \delta}{\pi} e^{(x/L) (G/2)} \nonumber \\
&\hspace{0 pt}\times \int_0^\infty \bigg[ \frac{4 y \, \D y}{(1+4 y^2)^2} \bigg( \sin\bigg( \frac{G y x}{L} \bigg) + 2 y \cos \bigg( \frac{G y x}{L} \bigg) \bigg) \nonumber \\
&\hspace{100 pt}\times \sin^2 \bigg( \frac{G c_s t}{L} \sqrt{y^2 + \frac{1}{4}} \bigg) \bigg] . \label{eqn:largeGSolution}
\end{align}
When $G$ becomes very large, the fluid becomes strongly rarefied and heated near $x = L$. When calculating the size of the temperature separation across the system, it makes more sense to compare the temperature at $x = 0$ with that at a scaled height $x = z_0 \log 10$. The integral in Eq.~(\ref{eqn:largeGSolution}) can be evaluated numerically, and the maximal difference between $T_1(t, z_0\log 10) / T_0$ and $T_1(t, 0) / T_0$ is about $0.49 \delta$ when $\gamma = 5/3$ (the minimum is about $-0.53 \delta$). Geyko and Fisch did not make an analytic prediction of this dependence, but they did investigate it numerically, and their simulations found $0.47 \delta$ for the maximum. 

Formally, the analytic calculations in this section are done in the limit of small $\delta$. It is natural to wonder how small $\delta$ has to be in order for the calculations to be accurate. The nonlinear fluid simulations shown in Figure~\ref{fig:simulations} shed some light on this point. When $\delta = 0.01$, the fluid simulations are almost indistinguishable from the analytic results. When $\delta$ is increased to $0.5$, the accuracy of the analytic results depends strongly on $G$. 

For $G = 0.1$ and $G = 1$, the $\delta = 0.5$ simulations are qualitatively very similar to the small-$\delta$ analytic results, except that the oscillations appear to take place at a higher frequency. This results from the temperature dependence of the system frequencies $\omega_n$. In Eq.~(\ref{eqn:omegaN}), these frequencies are written as functions of the pre-compression temperature $T_0$. However, physically, the system's frequency response after compression should scale with $T_i = (1 + \delta) T_0$ rather than $T_0$ (though the value of $T_0$ will determine which modes are excited). This distinction is not important when $\delta$ is small, but as $\delta$ grows larger it begins to matter. The simulations with $G = 0.1$ and $G = 1$ are dominated by the $n=1$ mode. If the frequency $\omega_1$ is evaluated at $T_i$ rather than $T_0$, $\omega_1$ increases by about 22\% when $G = 0.1$ or $1$. This is consistent with the higher-frequency $n=1$ modes observed in the simulations. 

However, when $\delta = 0.5$ and $G = 8$, the fluid simulations no longer resemble the small-$\delta$ calculations. This can be explained by the dependence of $T_1$ on $G$. $T_1$ depends nonlinearly on $G$, but in general $T_1$ grows larger as $G$ increases. As such, the $\delta$ that is required to keep $T_1 \ll T_0$ is smaller for larger values of $G$. For the simulations in Figure~\ref{fig:simulations}, $T - T_0 < T_0$ when $G = 0.1$ and $G = 1$, but when $G = 8$ and $\delta = 0.5$, there are regions with $T - T_0 > T_0$ and the perturbative model is no longer valid. 

\section{Arbitrary Compression Profiles} \label{sec:slowCompression}

The analysis in Section~\ref{sec:fastCompression} describes fast compression, so that the system starts out of equilibrium at $t = 0$ and is not driven after $t = 0$. It is possible to approach the case of more general heating profiles by instead allowing the system to start at equilibrium and imposing a time-dependent heat source. Suppose, to leading order, the heat source produces a spatially constant change in temperature. Then Eq.~(\ref{eqn:TLinear}) becomes 
\begin{align}
\frac{\partial T_1}{\partial t} = (\gamma-1) \frac{T_0}{n_0} \bigg( \frac{\partial n_1}{\partial t} - \frac{1}{z_0} \, v_1 n_0 \bigg) + \chi(t)
\end{align}
for some heating function $\chi(t)$. $f = v_1 e^{-x/2z_0}$ can be defined the same way, but its governing equation now depends on $\chi$:
\begin{align}
\frac{\partial^2 f}{\partial t^2} = c_s^2 \bigg( \frac{\partial^2 f}{\partial x^2} - \frac{1}{4 z_0^2} f \bigg) + \frac{\chi(t)}{m z_0} \, e^{x/2z_0} \, .
\end{align}
Define $\Phi_n(t)$ by 
\begin{align}
&\Phi_n(t) \doteq \frac{\omega_n}{T_0} \int_0^t \D t' \sin(\omega_n t') \int_0^{t'} \D t'' \, \chi(t'') \cos(\omega_n t'') \nonumber \\
&\hspace{16 pt}- \frac{\omega_n}{T_0} \int_0^t \D t' \cos(\omega_n t') \int_0^{t'} \D t'' \, \chi(t'') \sin(\omega_n t'') \, . \label{eqn:Phi}
\end{align}
In terms of $\Phi_n(t)$, the solution for $T_1$ is 
\begin{align}
&\frac{T_1}{T_0} = \int_0^t \frac{\chi(t') \D t'}{T_0} - \frac{\gamma-1}{\gamma} \big( 4 G \big) e^{(x/L)(G/2)} \nonumber \\
&\hspace{0 pt} \times \sum_{n=1}^\infty \bigg[ \frac{4 \pi n}{(G^2+4 \pi^2 n^2)^2} [1+(-1)^{n+1} e^{-G/2}] \nonumber \\
&\hspace{40 pt} \times \bigg( G \sin(k_n x) + 2 \pi n \cos(k_n x) \bigg) \Phi_n(t) \bigg]. \label{eqn:slowCompressionSolution}
\end{align}
Consider the case of steady heating for an interval $\tau$. Set
\begin{gather}
\chi(t) = \begin{cases}
\delta T_0 / \tau & 0 \leq t \leq \tau \\
0 & t<0 \text{, } t > \tau. 
\end{cases} \label{eqn:steadyPsi}
\end{gather}
Here, the parameter $\delta$ is analogous to the corresponding parameter in the fast-compression case. Using this choice of $\chi(t)$, 
\begin{align}
&\Phi_n(0 \leq t \leq \tau) = \bigg( \frac{t}{\tau} - \frac{\sin(\omega_n t)}{\omega_n \tau} \bigg) \delta
\end{align}
and 
\begin{align}
&\Phi(t > \tau) = \bigg( 1 + \frac{\sin(\omega_n(t-\tau))}{\omega_n \tau} - \frac{\sin(\omega_n t)}{\omega_n \tau} \bigg) \delta .\label{eqn:PhiAfterTau}
\end{align}
In the fast-compression limit where $\tau \rightarrow 0$, Eqs.~(\ref{eqn:slowCompressionSolution}) and (\ref{eqn:PhiAfterTau}) reduce to Eq.~(\ref{eqn:fastCompressionSolutionDimensionlessParameters}). On the other hand, in the limit of very slow compression, 
\begin{align}
&\lim_{\omega_n \tau \rightarrow \infty} \frac{T_1(t>\tau)}{T_0} = \delta - \frac{\gamma-1}{\gamma} \big( 4 G \delta \big) e^{(x/L)(G/2)} \nonumber \\
&\hspace{0 pt} \times \sum_{n=1}^\infty \bigg[ \frac{4 \pi n}{(G^2+4 \pi^2 n^2)^2} [1+(-1)^{n+1} e^{-G/2}] \nonumber \\
&\hspace{40 pt} \times \bigg( G \sin(k_n x) + 2 \pi n \cos(k_n x) \bigg) \bigg]. \label{eqn:verySlowCompressionSolution}
\end{align}
When $\omega_n \tau$ is large, the temperature gradient is not oscillatory. This is consistent with the intuition that a slowly driven system will remain close to force equilibrium. 
The temperature difference across the system can be written in closed form as 
\begin{align}
&\lim_{\omega_n \tau \rightarrow \infty} \frac{T_1(t > \tau, L) - T_1(t > \tau, 0)}{T_0} = \frac{\gamma - 1}{\gamma} \big(G \delta \big).
\end{align}
In the limit where $G \ll 1$, the temperature difference across the system for slow compression will be half of the maximal temperature difference for fast compression. This agrees exactly with the analytic result of Geyko and Fisch in that limit, though their simulations yielded a somewhat smaller coefficient. 

Of course, Eqs.~(\ref{eqn:Phi}) and (\ref{eqn:slowCompressionSolution}) make it clear that things can turn out quite differently if $\chi$ has a more complicated time dependence. It was already true in the simple case described by Eq.~(\ref{eqn:steadyPsi}) that a careful choice of $\tau$ could either suppress or enhance the oscillations associated with a particular mode number. If, for instance, $\chi$ itself were oscillatory, then particular modes could be driven or suppressed even more dramatically. Consider the oscillatory heating function 
\begin{gather}
\chi(t) = \delta \, \Omega \, T_0 \sin(\Omega t) 
\end{gather}
where $\Omega$ is some positive frequency. 
Heating of precisely this form may not necessarily be practically realizable, but it is an informative formal example. 
For this choice of $\chi$, 
\begin{gather}
\Phi_n(t) = \frac{[\omega_n^2 - \Omega^2 - \omega_n^2 \cos(\Omega t) + \Omega^2 \cos(\omega_n t) ] \delta}{\omega_n^2-\Omega^2} \, . 
\end{gather}
When the driving frequency is close to $\omega_n$, there is a secular term. To leading order in $\Omega - \omega_n$, 
\begin{gather}
\Phi_n(t) \rightarrow \bigg( 1 - \cos(\omega_n t) - \frac{\omega_n t}{2} \sin(\omega_n t) \bigg) \delta. 
\end{gather}
This holds even for higher-frequency oscillations whose role in the bulk behavior of the system would normally be small. Driving at one of the system's natural frequencies can produce temperature oscillations that (at least as far as the linear theory is concerned) can grow without bound. If the system is driven at $\omega_n$, the resonant oscillations will be associated with the corresponding spatial wavenumber $k_n$. All of this behavior is intuitive, if the system's response to $\chi(t)$ is understood in terms of the mode decomposition that comes naturally from the fluid picture. 

\section{Comparison of the fluid and Monte Carlo simulations}
\label{sec:two_codes}

As pointed out, the numerical results from the original paper on the piezothermal effect \cite{Geyko2014}, obtained via Monte Carlo simulations, are qualitatively similar to the ones obtained in the present work, yet deviate quantitatively in many cases. The main reason for this is the fact that the Monte Carlo code has intrinsic physical and numerical damping built in due to the finite mean free paths of the particles. To get a better understanding of this phenomenon, we briefly review the Monte Carlo code from the original paper.

The object of the simulations is a set of ideal particles that move in a one-dimensional box in a constant gravitational field $\textbf{g} = - g \hat{x}$. The box is considered infinite or periodic in the perpendicular directions $\hat{y}$ and $\hat{z}$, and of the length $L$ in the $\hat{x}$ direction. Particle velocities, however, have all three components ($v_x$, $v_y$, and $v_z$) for the sake of preserving the proper value of the adiabatic gas constant $\gamma=5/3$. A particle's motion is exactly integrated for every time step $\delta t$, and takes into account the possibility of multiple particle-wall collisions on the box floor.

A non-interacting ensemble of particles does not represent a fluid-like motion. Instead, it will produce complex but uncorrelated behavior, like the density waves described in \cite{Kolmes2016}. In order to make the system behave like a fluid, particle collisions are added. In the code, only binary elastic collisions are considered, such that energy, momentum, and angular momentum are conserved up to machine precision for each individual collision and, as a result, for the whole system. The main problem of such a collision operator is that any two particles are never located at the same point in space. In principle, a given pair of particles can be tracked and the time of the true collision can be found, yet this is too complicated if all the particles are required to collide every time step. Thus, some nearly located particles are picked for each collision. The domain is divided in the $\hat{x}$ direction into a number of cells, each of the same length $L_c$ for simplicity. Since the particles are not at exactly the same point, the collision should be acting along the direction $\hat{\ell}$ connecting the centers of the two particles, otherwise the angular momentum will not be conserved. One can think about this type of collision as an instantaneous force acting between the two particles, like gravitational attraction. This force should change somehow the projections of particle velocities $v_{1\ell}$ and $v_{2\ell}$ in such a way that the total kinetic energy and momentum are conserved. For identical particles, it is done by exchanging their velocity projections: $v_{1\ell} \to v_{2\ell}$ and $v_{2\ell} \to v_{1\ell}$. Since the two particles are picked at random inside a cell, the distance $d$ between them is of the order of $L_c$. The angle between the direction $\hat{\ell}$ and $\hat{x}$ is $\theta$, and it is picked at random but is typically about $\theta\approx \pi/3$ or similar, because the perpendicular displacement is picked uniformly in both directions from $-L_c$ to $L_c$.

This collision operator exactly conserves energy, momentum and angular momentum, but suffers from numerical heat and momentum transfer due to finite cell size effects. This can be understood in the following way: imagine the cell size is equal to the box height, and a hot population of the particles is sitting at the bottom. In this case, the numerical thermalization would occur instantly, and the particles on the top would get hot even faster than a sound wave can travel across the domain.

To be more specific, consider two particles inside a cell located at coordinates $x_1$ and $x_2$, respectively. For highly collisional gas, which is of interest here, a Maxwellian distribution can be assumed, with temperature $T(x)$, mean velocity $u(x) \hat x$, and density $n(x)$. As a collision occurs, an instantaneous transfer of the momentum from the second particle to the first one can be written as
\begin{gather}
\frac{\Delta \textbf{p}}{m} = \int d^3v_1 f_1(\textbf{v}_1, x_1) \int d^3v_2 f_2(\textbf{v}_2, x_2) [\tilde{\textbf{v}}_2 - \tilde{\textbf{v}}_1] ,
\end{gather}
where $\tilde{\textbf{v}}$ is a projection of the velocity to the $\hat{\ell}$ direction $\tilde{\textbf{v}}=\hat{\ell}(\hat{\ell}\cdot\textbf{v})$. Integrals with respect to $v_y$ and $v_z$ vanish, because the integrated function is antisymmetric, and the integral with respect to $v_x$ yields 
\begin{gather}\label{eq:mom_1}
\Delta \textbf{p} = m \hat{\ell} \ell_x \left( u(x_2)-u(x_1)\right),
\end{gather}
where only $\Delta p_x$ is of interest since the other two components vanish, as an averaging over $\hat{\ell}$ is performed, thus,
\begin{gather}\label{eq:mom_2}
\Delta p = \Delta p_x = m \cos^2(\theta)\left( u(x_2)-u(x_1)\right).
\end{gather}

For a particle at a given position $\bar{x}$ inside the cell ($\bar{x}=0$ at the center of the cell), the total momentum transfer from all the particles around is found as a mass weighed integral over all the cell of Eq.~(\ref{eq:mom_2}), where 
%$\cos^2(\theta)$ can be approximated as some constant number $c_1$ and where 
density and velocity are Taylor expanded around the cell-center point $x_c$. This integral should be also multiplied by a collision rate parameter $R_c$, which is proportional to the number of collisions occurred in the given cell each time step.
\begin{align}\label{eq:mom_tot}
&\Delta p_\text{tot} = m R_c \int\limits_{-L_c/2}^{L_c/2} \cos^2 \theta \left[\left(n_c + n'\xi + \frac{n''}{2}\xi^2\right) \right. \nonumber \\ 
&\hspace{70 pt}\left. \cdot \left( u'(\xi-\bar{x}) +\frac{u''}{2}(\xi^2 -\bar{x}^2) \right) \right]d\xi.
\end{align}
The result of expression~(\ref{eq:mom_tot}) depends on the value of $\bar{x}$, however for any $\bar{x}$ there always present a term proportional to $m R_c n_c u'' L_c^3$. Notice that $n_c L_c \approx N_p$, where $N_p$ is the number of particles in the cell, and the momentum transfer found in Eq.~(\ref{eq:mom_tot}) happens in a time step $\delta t$. Therefore, there is a momentum transfer term with 
\begin{gather}
\frac{\partial p}{\partial t} \propto \frac{m R_c N_p L_c^2}{\delta t}\frac{\partial^2 u}{\partial x^2}, 
\end{gather}
and Eq.~(\ref{eqn:nonlinearV}) then reads as
\begin{gather}
m n \bigg( \frac{\partial v}{\partial t} + v \frac{\partial v}{\partial x} \bigg) = - \frac{\partial (nT)}{\partial x} - m n g + \nu m n \frac{\partial^2 u}{\partial x^2},
\end{gather}
where $\nu$ is the derived numerical viscosity with $\nu \propto R_c L_c^2/\delta t$. The derivation of numerical heat conductivity is very similar to the one for viscosity, and therefore is omitted here. % changed \nu = to \nu \propto

\begin{figure*}
	\centering
	\includegraphics[width=\linewidth]{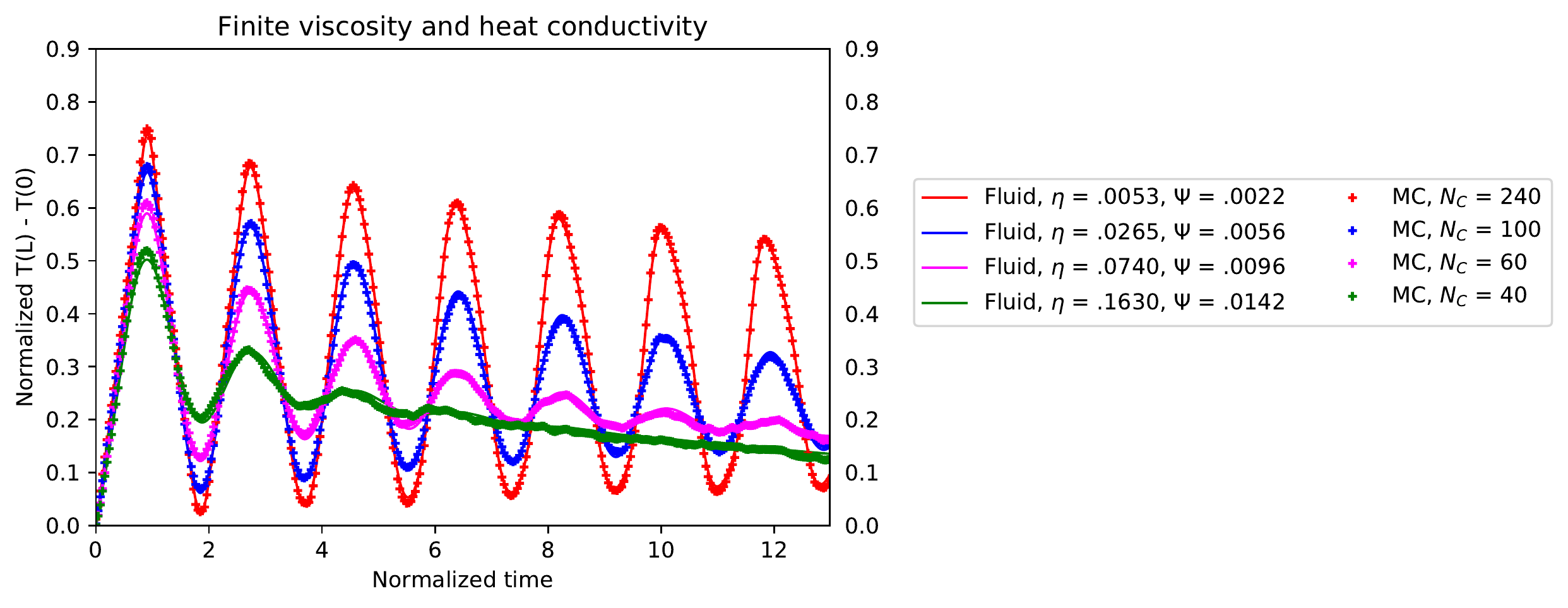}
	\caption{Evolution of the temperature difference $T(L)-T(0)$ normalized to $G\delta T_0$ in a series of Monte Carlo and fluid simulations. The grid parameter $N_c$ is varied for the Monte Carlo simulations. The viscosity $\eta$ and heat conductivity $\Psi$ are varied for the fluid simulations. All other code parameters are fixed. The listed values of $\eta$ and $\Psi$ are normalized to the product of the system height and the sound speed.}
	\label{fig:osc_NC}
\end{figure*}

Apart from numerical viscosity and heat conductivity, driven mainly by a finite cell size, there is a physical mechanism of heat conductivity due to finite particle mean free path. The last is determined be the collision rate $R_c$, the time step $\delta t$, and the mean particle velocity $v_t$ and does not depend on the cell size. Indeed, consider a generalized version of Eq.~(\ref{eqn:nonlinearT}) with heat transfer term included in it
\begin{gather}
\frac{n}{\gamma-1}\left(\frac{\partial T}{\partial t} + v\frac{\partial T}{\partial x}\right) + nT\frac{\partial v}{\partial x} = \frac{\partial}{\partial x}\left(\Psi \frac{\partial T}{\partial x}\right).\label{eqn:energy2}
\end{gather}
Here, $\Psi$ is the heat conductivity coefficient, given in terms of the mean free path $\lambda_\text{mfp}$ as $\Psi \approx n \lambda_\text{mfp} v_t c_v/3$. Eq.~(\ref{eqn:energy2}) reduces to Eq.~(\ref{eqn:nonlinearT}) if $\Psi = 0$. When $\Psi > 0$, heat diffusion leads to wave dissipation and system equilibration. 

%\begin{align}\label{eq:energy_1}
% &\frac{\partial}{\partial t}\left( \frac{mnv^2}{2}+c_v nT\right) \nonumber \\
% &\hspace{30 pt}+ \frac{\partial}{\partial x} \left(nv\left[\frac{mv^2}{2}+c_pT\right] -\Psi \frac{\partial T}{\partial x} \right) = \dot{Q}.
%\end{align}
%Here, $\dot{Q}$ is the energy deposition rate, equal to $-mgvn$ for this particular problem, $\Psi$ is the heat conductivity coefficient, given in terms of the mean free path $\lambda_\text{mfp}$ as $\Psi \approx n \lambda_\text{mfp} v_t c_v/3$. Eq.~(\ref{eq:energy_1}) reduces to Eq.~(\ref{eqn:nonlinearT}) if $\Psi=0$, otherwise, it takes a form % heat -> energy [deposition]
%\begin{gather}
%\frac{n}{\gamma-1}\left(\frac{\partial T}{\partial t} + v\frac{\partial T}{\partial x}\right) + nT\frac{\partial v}{\partial x} = \frac{\partial}{\partial x}\left(\Psi \frac{\partial T}{\partial x}\right),
%\end{gather}
%where the last term represents heat diffusion and leads to the wave dissipation and system equilibration.

Notice that the aforementioned arguments are not a rigorous derivation of the numerical viscosity and heat conductivity in the Monte Carlo code. They can only provide some insights on why Monte Carlo simulations sometimes produce different results. However, even such a simplified picture is enough to explain, for example, why the piezothermal coefficient 
\begin{gather}
\kappa \doteq \frac{T_1(L/c_s,L)-T_1(L/c_s,0)}{G\delta T_0}
\end{gather} was 0.64 instead of 0.8 (see Eq.~(\ref{eq:kappa_limit})) in the numerical results from the original paper. In particular, we are interested in how $\kappa$ depends on the length $L_c$, which was described by a parameter $N_c$ in the code, where $N_c L_c = 1$.

Figure~\ref{fig:osc_NC} shows how the piezothermal temperature difference evolves as a function of time in a series of simulations using two different codes: one performing Monte Carlo simulations and the other performing fluid simulations. The Monte Carlo simulations, denoted by plus marks, show the temperature difference for four different values of $N_c$, while all other parameters of the code were fixed, namely, $\delta t = 0.001$, $T_0 = 0.3698$, $R_c = 10$ (collisions per particle per cell), $G = 1.352$, $\delta T_0 = 0.0518$. 
%Fig.~(\ref{fig:osc_NC}) shows how piezothermal temperature difference evolve as a function of time for four different values of $N_c$, while all other parameters of the code were fixed, namely, $\delta t = 0.001$, $T_0 = 0.3698$, $R_c = 10$ (collisions per particle in a cell), $G=1.352$, $\delta T_0 = 0.0518$. 
Only for $N_c=240$ the first peak of the oscillations is sufficiently close to the predicted value 0.8, yet the oscillations nevertheless slowly damp in time. For low values of $N_c$ fluid oscillations are very quickly damped, and the system decays to a new equilibrium. 

The solid lines in Figure~\ref{fig:osc_NC} show a corresponding series of fluid simulations. In these simulations, the field strength parameter $G$ and the heating parameter $\delta$ are chosen to match the values in the Monte Carlo simulations. Each of these fluid simulations includes a spatially constant viscosity $\eta$ and heat conductivity $\Psi$. Of course, discretization error is not a phenomenon unique to Monte Carlo algorithms. Fluid simulations also have finite-grid-size effects. The fluid simulations shown here use sufficiently fine-grained grids that these errors are negligible compared to the corresponding effects in the Monte Carlo code (in this example, the fluid simulations used 128 cells). 

Both the Monte Carlo simulations and the fluid simulations show oscillations that are ``lopsided," in the sense that they are asymmetric about their extrema. The asymmetry is most apparent in the $N_c = 240$ case. This results from the same nonlinearity discussed at the end of Section~\ref{sec:fastCompression}, in which $\delta$ and $G$ are large enough for the oscillations not to be small perturbations. It is worth noting that these asymmetric oscillations still appear even in fluid simulations without any viscosity or heat conductivity (not shown in Figure~\ref{fig:osc_NC}). 

In any case, there are two major conclusions to be drawn from the comparison in Figure~\ref{fig:osc_NC}. First, the finite-cell-size effects seen in the Monte Carlo simulations appear to be equivalent to an effective viscosity and heat conductivity. Second, the effective viscosity and heat conductivity become small when $N_c$ is large. 

\section{Discussion and Conclusions}

Using a fluid model, we have derived analytic expressions for the temperature gradients of the piezothermal effect as they evolve in time. The fluid solutions recover the original analytic model's predictions for $G \ll 1$ and they make it possible to make predictions when $G$ is not small. Similarly, they recover the original model's qualitative predictions for very slow and very fast compression while also handling more general compression profiles, including compression that is not constant in time and compression that is neither very fast nor very slow. The analytic solutions to the fluid equations are in very good agreement with fluid simulations performed using the SNeuT fluid code. 

There are places where the results from fluid models disagree quantitatively with some of the numerical results from the original paper. 
The comparison between the present fluid and the original Monte Carlo simulations provides some explanation for why the previous results were different, and what can be done in order to improve them in the Monte Carlo model. In general, a small time step and a very large number of cells are required in order to sufficiently suppress numerical and physical heat diffusion and viscosity in the Monte Carlo simulations. That brings extra complication for the total number of particles in the system, as the number of particles in a cell should be large enough to mitigate statistical noise. However, there is evidence that (in the appropriate limit) the Monte Carlo simulations converge to results that agree with the fluid model. 

The fluid model used in this paper makes assumptions. The strict timescale ordering means that viscosity and heat conductivity are neglected (with the exception of the simulations used to produce Figure~\ref{fig:osc_NC}, which included both), though the calculation in Section~\ref{sec:slowCompression} makes it possible to relax the requirement for an ordering between the compression timescale $\tau_E$ and the sound timescale $\tau_s$. The analytic calculations presented here use linearized fluid equations; they become invalid when the compression parameter $\delta$ is large. However, these assumptions were also necessary for the model used in the original paper. 

The mode structure of the analytic solutions helps to provide intuition for the behavior of the piezothermal effect. The critical dependence of the effect on the field-strength parameter $G$ can be explained by the mode structure: as $G$ increases, modes other than $n=1$ become important when $n \lesssim G / 2 \pi$. When $G$ is small, the piezothermal effect is dominated by a single frequency and a single wavenumber; when $G$ is large, many frequencies and wavenumbers contribute, and the oscillations can become much more complicated. 

The characteristic frequencies $\omega_n$ are closely related to the Brunt-V\"ais\"al\"a frequency, which is important in a variety of geophysical, astrophysical, oceanographic, and atmospheric contexts \cite{Brunt1927, Durran1982, Emery1984, Brassard1991}. Brunt-V\"ais\"al\"a oscillations occur when a fluid element is displaced within a stratified background. 
For a parcel of air displaced in a dry, isothermal atmosphere, the Brunt-V\"ais\"al\"a frequency can be written as \cite{Brunt1927}
\begin{gather}
\omega_\text{BV} = \sqrt{ \frac{g \Gamma_d}{T} } = \sqrt{ \frac{g^2}{c_p T} } = \frac{2 \omega_0}{\sqrt{\gamma m c_p}} \, ,
\end{gather}
where $\Gamma_d$ is the dry adiabatic lapse rate and $c_p$ is the specific heat capacity. 

The scenario being considered here is not quite identical to the prototypical Brunt-V\"ais\"al\"a buoyancy oscillation; for one thing, the entire system is displaced, rather than a small fluid element within the system. However, the oscillations associated with the piezothermal effect can be understood as a spectrum of buoyancy oscillations which are closely related to Brunt-V\"ais\"al\"a oscillations. 

\section*{Acknowledgements}

VIG was supported under the auspices of the U.S. Department of Energy by Lawrence Livermore National Laboratory under Contract DE-AC52-07NA27344. EJK and NJF were supported by NSF PHY-1506122 and NNSA 83228-10966 [Prime No. DOE (NNSA) DE-NA0003764]. The SNeuT simulation code uses the CVODE package, an open source software package which is part of Lawrence Livermore National Laboratory's SUNDIALS suite. SNeuT is a fork of the MITNS plasma transport code \cite{KolmesMITNSarXiv}. Authors are thankful to Eric Emdee, Mike Mlodik, and Jace Waybright for fruitful discussions, and to Ian Ochs for fruitful discussions and for involvement in code development. 

\bibliographystyle{apsrev4-1} 
\bibliography{../../../Master}

\providecommand{\noopsort}[1]{}\providecommand{\singleletter}[1]{#1}%
\begin{thebibliography}{21}%
\makeatletter
\providecommand \@ifxundefined [1]{%
 \@ifx{#1\undefined}
}%
\providecommand \@ifnum [1]{%
 \ifnum #1\expandafter \@firstoftwo
 \else \expandafter \@secondoftwo
 \fi
}%
\providecommand \@ifx [1]{%
 \ifx #1\expandafter \@firstoftwo
 \else \expandafter \@secondoftwo
 \fi
}%
\providecommand \natexlab [1]{#1}%
\providecommand \enquote  [1]{``#1''}%
\providecommand \bibnamefont  [1]{#1}%
\providecommand \bibfnamefont [1]{#1}%
\providecommand \citenamefont [1]{#1}%
\providecommand \href@noop [0]{\@secondoftwo}%
\providecommand \href [0]{\begingroup \@sanitize@url \@href}%
\providecommand \@href[1]{\@@startlink{#1}\@@href}%
\providecommand \@@href[1]{\endgroup#1\@@endlink}%
\providecommand \@sanitize@url [0]{\catcode `\\12\catcode `\$12\catcode
  `\&12\catcode `\#12\catcode `\^12\catcode `\_12\catcode `\%12\relax}%
\providecommand \@@startlink[1]{}%
\providecommand \@@endlink[0]{}%
\providecommand \url  [0]{\begingroup\@sanitize@url \@url }%
\providecommand \@url [1]{\endgroup\@href {#1}{\urlprefix }}%
\providecommand \urlprefix  [0]{URL }%
\providecommand \Eprint [0]{\href }%
\providecommand \doibase [0]{http://dx.doi.org/}%
\providecommand \selectlanguage [0]{\@gobble}%
\providecommand \bibinfo  [0]{\@secondoftwo}%
\providecommand \bibfield  [0]{\@secondoftwo}%
\providecommand \translation [1]{[#1]}%
\providecommand \BibitemOpen [0]{}%
\providecommand \bibitemStop [0]{}%
\providecommand \bibitemNoStop [0]{.\EOS\space}%
\providecommand \EOS [0]{\spacefactor3000\relax}%
\providecommand \BibitemShut  [1]{\csname bibitem#1\endcsname}%
\let\auto@bib@innerbib\@empty
%</preamble>
\bibitem [{\citenamefont {Geyko}\ and\ \citenamefont
  {Fisch}(2016)}]{Geyko2016}%
  \BibitemOpen
  \bibfield  {author} {\bibinfo {author} {\bibfnamefont {V.~I.}\ \bibnamefont
  {Geyko}}\ and\ \bibinfo {author} {\bibfnamefont {N.~J.}\ \bibnamefont
  {Fisch}},\ }\href {\doibase 10.1103/PhysRevE.94.042113} {\bibfield  {journal}
  {\bibinfo  {journal} {Phys. Rev. E}\ }\textbf {\bibinfo {volume} {94}},\
  \bibinfo {pages} {042113} (\bibinfo {year} {2016})}\BibitemShut {NoStop}%
\bibitem [{\citenamefont {Geyko}\ and\ \citenamefont
  {Fisch}(2013)}]{Geyko2013}%
  \BibitemOpen
  \bibfield  {author} {\bibinfo {author} {\bibfnamefont {V.~I.}\ \bibnamefont
  {Geyko}}\ and\ \bibinfo {author} {\bibfnamefont {N.~J.}\ \bibnamefont
  {Fisch}},\ }\href {\doibase 10.1103/PhysRevLett.110.150604} {\bibfield
  {journal} {\bibinfo  {journal} {Phys. Rev. Lett.}\ }\textbf {\bibinfo
  {volume} {110}},\ \bibinfo {pages} {150604} (\bibinfo {year}
  {2013})}\BibitemShut {NoStop}%
\bibitem [{\citenamefont {Geyko}\ and\ \citenamefont
  {Fisch}(2017)}]{Geyko2017}%
  \BibitemOpen
  \bibfield  {author} {\bibinfo {author} {\bibfnamefont {V.~I.}\ \bibnamefont
  {Geyko}}\ and\ \bibinfo {author} {\bibfnamefont {N.~J.}\ \bibnamefont
  {Fisch}},\ }\href {\doibase 10.1063/1.4975651} {\bibfield  {journal}
  {\bibinfo  {journal} {Phys. Plasmas}\ }\textbf {\bibinfo {volume} {24}},\
  \bibinfo {pages} {022113} (\bibinfo {year} {2017})}\BibitemShut {NoStop}%
\bibitem [{\citenamefont {Geyko}\ and\ \citenamefont
  {Fisch}(2014)}]{Geyko2014}%
  \BibitemOpen
  \bibfield  {author} {\bibinfo {author} {\bibfnamefont {V.~I.}\ \bibnamefont
  {Geyko}}\ and\ \bibinfo {author} {\bibfnamefont {N.~J.}\ \bibnamefont
  {Fisch}},\ }\href {\doibase 10.1103/PhysRevE.90.022139} {\bibfield  {journal}
  {\bibinfo  {journal} {Phys. Rev. E}\ }\textbf {\bibinfo {volume} {90}},\
  \bibinfo {pages} {022139} (\bibinfo {year} {2014})}\BibitemShut {NoStop}%
\bibitem [{\citenamefont {Ranque}(1933)}]{Ranque1933}%
  \BibitemOpen
  \bibfield  {author} {\bibinfo {author} {\bibfnamefont {G.}~\bibnamefont
  {Ranque}},\ }\href@noop {} {\bibfield  {journal} {\bibinfo  {journal} {J.
  Phys. Rad.}\ }\textbf {\bibinfo {volume} {7 (4)}},\ \bibinfo {pages} {112}
  (\bibinfo {year} {1933})}\BibitemShut {NoStop}%
\bibitem [{\citenamefont {Hilsch}(1947)}]{Hilsch1947}%
  \BibitemOpen
  \bibfield  {author} {\bibinfo {author} {\bibfnamefont {R.}~\bibnamefont
  {Hilsch}},\ }\href {\doibase 10.1063/1.1740893} {\bibfield  {journal}
  {\bibinfo  {journal} {Rev. Sci. Instrum.}\ }\textbf {\bibinfo {volume}
  {18}},\ \bibinfo {pages} {108} (\bibinfo {year} {1947})}\BibitemShut
  {NoStop}%
\bibitem [{\citenamefont {Kassner}\ and\ \citenamefont
  {Knoernschild}(1948)}]{Kassner1948}%
  \BibitemOpen
  \bibfield  {author} {\bibinfo {author} {\bibfnamefont {R.}~\bibnamefont
  {Kassner}}\ and\ \bibinfo {author} {\bibfnamefont {E.}~\bibnamefont
  {Knoernschild}},\ }\href@noop {} {\emph {\bibinfo {title} {Friction Laws and
  Energy Transfer in Circular Flow}}},\ \bibinfo {type} {Tech. Rep.}\ \bibinfo
  {number} {F-TR-2198-ND}\ (\bibinfo  {institution} {Wright-Patterson Air Force
  Base},\ \bibinfo {year} {1948})\BibitemShut {NoStop}%
\bibitem [{\citenamefont {Ahlborn}\ and\ \citenamefont
  {Groves}(1997)}]{Ahlborn1997}%
  \BibitemOpen
  \bibfield  {author} {\bibinfo {author} {\bibfnamefont {B.}~\bibnamefont
  {Ahlborn}}\ and\ \bibinfo {author} {\bibfnamefont {S.}~\bibnamefont
  {Groves}},\ }\href {\doibase 10.1016/S0169-5983(97)00003-8} {\bibfield
  {journal} {\bibinfo  {journal} {Fluid Dyn. Research}\ }\textbf {\bibinfo
  {volume} {21}},\ \bibinfo {pages} {73} (\bibinfo {year} {1997})}\BibitemShut
  {NoStop}%
\bibitem [{\citenamefont {Ahlborn}\ \emph {et~al.}(1998)\citenamefont
  {Ahlborn}, \citenamefont {Keller},\ and\ \citenamefont
  {Rebhan}}]{Ahlborn1998}%
  \BibitemOpen
  \bibfield  {author} {\bibinfo {author} {\bibfnamefont {B.~K.}\ \bibnamefont
  {Ahlborn}}, \bibinfo {author} {\bibfnamefont {J.~U.}\ \bibnamefont {Keller}},
  \ and\ \bibinfo {author} {\bibfnamefont {E.}~\bibnamefont {Rebhan}},\ }\href
  {\doibase 10.1515/jnet.1998.23.2.159} {\bibfield  {journal} {\bibinfo
  {journal} {J. Non-Equilib. Thermodyn.}\ }\textbf {\bibinfo {volume} {23}},\
  \bibinfo {pages} {159} (\bibinfo {year} {1998})}\BibitemShut {NoStop}%
\bibitem [{\citenamefont {Ahlborn}\ and\ \citenamefont
  {Gordon}(2000)}]{Ahlborn2000}%
  \BibitemOpen
  \bibfield  {author} {\bibinfo {author} {\bibfnamefont {B.~K.}\ \bibnamefont
  {Ahlborn}}\ and\ \bibinfo {author} {\bibfnamefont {J.~M.}\ \bibnamefont
  {Gordon}},\ }\href {\doibase 10.1063/1.1289524} {\bibfield  {journal}
  {\bibinfo  {journal} {J. Appl. Phys.}\ }\textbf {\bibinfo {volume} {88}},\
  \bibinfo {pages} {3645} (\bibinfo {year} {2000})}\BibitemShut {NoStop}%
\bibitem [{\citenamefont {Liew}\ \emph {et~al.}(2012)\citenamefont {Liew},
  \citenamefont {Zeegers}, \citenamefont {Kuerten},\ and\ \citenamefont
  {Michalek}}]{Liew2012}%
  \BibitemOpen
  \bibfield  {author} {\bibinfo {author} {\bibfnamefont {R.}~\bibnamefont
  {Liew}}, \bibinfo {author} {\bibfnamefont {J.~C.~H.}\ \bibnamefont
  {Zeegers}}, \bibinfo {author} {\bibfnamefont {J.~G.~M.}\ \bibnamefont
  {Kuerten}}, \ and\ \bibinfo {author} {\bibfnamefont {W.~R.}\ \bibnamefont
  {Michalek}},\ }\href {\doibase 10.1103/PhysRevLett.109.054503} {\bibfield
  {journal} {\bibinfo  {journal} {Phys. Rev. Lett.}\ }\textbf {\bibinfo
  {volume} {109}},\ \bibinfo {pages} {054503} (\bibinfo {year}
  {2012})}\BibitemShut {NoStop}%
\bibitem [{\citenamefont {Kolmes}\ \emph {et~al.}(2017)\citenamefont {Kolmes},
  \citenamefont {Geyko},\ and\ \citenamefont {Fisch}}]{Kolmes2017}%
  \BibitemOpen
  \bibfield  {author} {\bibinfo {author} {\bibfnamefont {E.~J.}\ \bibnamefont
  {Kolmes}}, \bibinfo {author} {\bibfnamefont {V.~I.}\ \bibnamefont {Geyko}}, \
  and\ \bibinfo {author} {\bibfnamefont {N.~J.}\ \bibnamefont {Fisch}},\ }\href
  {\doibase 10.1016/j.ijheatmasstransfer.2016.11.072} {\bibfield  {journal}
  {\bibinfo  {journal} {Int. J. Heat Mass Transfer}\ }\textbf {\bibinfo
  {volume} {107}},\ \bibinfo {pages} {771} (\bibinfo {year}
  {2017})}\BibitemShut {NoStop}%
\bibitem [{\citenamefont {Davidovits}\ and\ \citenamefont
  {Fisch}(2016)}]{Davidovits2016}%
  \BibitemOpen
  \bibfield  {author} {\bibinfo {author} {\bibfnamefont {S.}~\bibnamefont
  {Davidovits}}\ and\ \bibinfo {author} {\bibfnamefont {N.~J.}\ \bibnamefont
  {Fisch}},\ }\href {\doibase 10.1103/PhysRevLett.116.105004} {\bibfield
  {journal} {\bibinfo  {journal} {Phys. Rev. Lett.}\ }\textbf {\bibinfo
  {volume} {116}},\ \bibinfo {pages} {105004} (\bibinfo {year}
  {2016})}\BibitemShut {NoStop}%
\bibitem [{\citenamefont {Brunt}(1927)}]{Brunt1927}%
  \BibitemOpen
  \bibfield  {author} {\bibinfo {author} {\bibfnamefont {D.}~\bibnamefont
  {Brunt}},\ }\href {\doibase 10.1002/qj.49705322103} {\bibfield  {journal}
  {\bibinfo  {journal} {Q. J. Royal Meteorol. Soc.}\ }\textbf {\bibinfo
  {volume} {53}},\ \bibinfo {pages} {30} (\bibinfo {year} {1927})}\BibitemShut
  {NoStop}%
\bibitem [{\citenamefont {Durran}\ and\ \citenamefont
  {Klemp}(1982)}]{Durran1982}%
  \BibitemOpen
  \bibfield  {author} {\bibinfo {author} {\bibfnamefont {D.~R.}\ \bibnamefont
  {Durran}}\ and\ \bibinfo {author} {\bibfnamefont {J.~B.}\ \bibnamefont
  {Klemp}},\ }\href {\doibase 10.1175/1520-0469(1982)039<2152:OTEOMO>2.0.CO;2}
  {\bibfield  {journal} {\bibinfo  {journal} {J. Atmospheric Sci.}\ }\textbf
  {\bibinfo {volume} {39}},\ \bibinfo {pages} {2152} (\bibinfo {year}
  {1982})}\BibitemShut {NoStop}%
\bibitem [{\citenamefont {Emery}\ \emph {et~al.}(1984)\citenamefont {Emery},
  \citenamefont {Lee},\ and\ \citenamefont {Magaard}}]{Emery1984}%
  \BibitemOpen
  \bibfield  {author} {\bibinfo {author} {\bibfnamefont {W.~J.}\ \bibnamefont
  {Emery}}, \bibinfo {author} {\bibfnamefont {W.~G.}\ \bibnamefont {Lee}}, \
  and\ \bibinfo {author} {\bibfnamefont {L.}~\bibnamefont {Magaard}},\ }\href
  {\doibase 10.1175/1520-0485(1984)014<0294:GASDOB>2.0.CO;2} {\bibfield
  {journal} {\bibinfo  {journal} {J. Phys. Oceanogr.}\ }\textbf {\bibinfo
  {volume} {14}},\ \bibinfo {pages} {294} (\bibinfo {year} {1984})}\BibitemShut
  {NoStop}%
\bibitem [{\citenamefont {Brassard}\ \emph {et~al.}(1991)\citenamefont
  {Brassard}, \citenamefont {Fontaine}, \citenamefont {Wesemael}, \citenamefont
  {Kawaler},\ and\ \citenamefont {Tassoul}}]{Brassard1991}%
  \BibitemOpen
  \bibfield  {author} {\bibinfo {author} {\bibfnamefont {P.}~\bibnamefont
  {Brassard}}, \bibinfo {author} {\bibfnamefont {G.}~\bibnamefont {Fontaine}},
  \bibinfo {author} {\bibfnamefont {F.}~\bibnamefont {Wesemael}}, \bibinfo
  {author} {\bibfnamefont {S.~D.}\ \bibnamefont {Kawaler}}, \ and\ \bibinfo
  {author} {\bibfnamefont {M.}~\bibnamefont {Tassoul}},\ }\href {\doibase
  10.1086/169655} {\bibfield  {journal} {\bibinfo  {journal} {Astrophys. J.}\
  }\textbf {\bibinfo {volume} {367}},\ \bibinfo {pages} {601} (\bibinfo {year}
  {1991})}\BibitemShut {NoStop}%
\bibitem [{\citenamefont {Kolmes}\ \emph {et~al.}(2016)\citenamefont {Kolmes},
  \citenamefont {Geyko},\ and\ \citenamefont {Fisch}}]{Kolmes2016}%
  \BibitemOpen
  \bibfield  {author} {\bibinfo {author} {\bibfnamefont {E.~J.}\ \bibnamefont
  {Kolmes}}, \bibinfo {author} {\bibfnamefont {V.~I.}\ \bibnamefont {Geyko}}, \
  and\ \bibinfo {author} {\bibfnamefont {N.~J.}\ \bibnamefont {Fisch}},\ }\href
  {\doibase 10.1016/j.physleta.2016.07.015} {\bibfield  {journal} {\bibinfo
  {journal} {Phys. Lett. A}\ }\textbf {\bibinfo {volume} {380}},\ \bibinfo
  {pages} {3061} (\bibinfo {year} {2016})}\BibitemShut {NoStop}%
\bibitem [{\citenamefont {Hindmarsh}\ \emph {et~al.}(2005)\citenamefont
  {Hindmarsh}, \citenamefont {Brown}, \citenamefont {Grant}, \citenamefont
  {Lee}, \citenamefont {Serban}, \citenamefont {Shumaker},\ and\ \citenamefont
  {Woodward}}]{Hindmarsh2005}%
  \BibitemOpen
  \bibfield  {author} {\bibinfo {author} {\bibfnamefont {A.~C.}\ \bibnamefont
  {Hindmarsh}}, \bibinfo {author} {\bibfnamefont {P.~N.}\ \bibnamefont
  {Brown}}, \bibinfo {author} {\bibfnamefont {K.~E.}\ \bibnamefont {Grant}},
  \bibinfo {author} {\bibfnamefont {S.~L.}\ \bibnamefont {Lee}}, \bibinfo
  {author} {\bibfnamefont {R.}~\bibnamefont {Serban}}, \bibinfo {author}
  {\bibfnamefont {D.~E.}\ \bibnamefont {Shumaker}}, \ and\ \bibinfo {author}
  {\bibfnamefont {C.~S.}\ \bibnamefont {Woodward}},\ }\href {\doibase
  10.1145/1089014.1089020} {\bibfield  {journal} {\bibinfo  {journal} {ACM
  Trans. Math. Softw.}\ }\textbf {\bibinfo {volume} {31}},\ \bibinfo {pages}
  {363} (\bibinfo {year} {2005})}\BibitemShut {NoStop}%
\bibitem [{\citenamefont {Cohen}\ \emph {et~al.}(1996)\citenamefont {Cohen},
  \citenamefont {Hindmarsh},\ and\ \citenamefont {Dubois}}]{Cohen1996}%
  \BibitemOpen
  \bibfield  {author} {\bibinfo {author} {\bibfnamefont {S.~D.}\ \bibnamefont
  {Cohen}}, \bibinfo {author} {\bibfnamefont {A.~C.}\ \bibnamefont
  {Hindmarsh}}, \ and\ \bibinfo {author} {\bibfnamefont {P.~F.}\ \bibnamefont
  {Dubois}},\ }\href {\doibase 10.1063/1.4822377} {\bibfield  {journal}
  {\bibinfo  {journal} {Comput. Phys.}\ }\textbf {\bibinfo {volume} {10}},\
  \bibinfo {pages} {138} (\bibinfo {year} {1996})}\BibitemShut {NoStop}%
\bibitem [{\citenamefont {Kolmes}\ \emph {et~al.}(2020)\citenamefont {Kolmes},
  \citenamefont {Ochs},\ and\ \citenamefont {Fisch}}]{KolmesMITNSarXiv}%
  \BibitemOpen
  \bibfield  {author} {\bibinfo {author} {\bibfnamefont {E.~J.}\ \bibnamefont
  {Kolmes}}, \bibinfo {author} {\bibfnamefont {I.~E.}\ \bibnamefont {Ochs}}, \
  and\ \bibinfo {author} {\bibfnamefont {N.~J.}\ \bibnamefont {Fisch}},\
  }\href@noop {} {}\bibinfo {howpublished} {arXiv:2002.09110} (\bibinfo {year}
  {2020})\BibitemShut {NoStop}%
\end{thebibliography}%

\end{document}